\documentstyle[12pt]{article}

\begin{document}

\thispagestyle{empty}

\vspace*{-2cm}
\begin{flushright}
CERN-TH/99-44\\hep-ph/9902455
\end{flushright}
\vspace*{1.5cm}

\begin{center}
{\bf\Large{Pion Fluctuations near the QCD Critical Point}}\\
\vspace{1.5 cm}
N.G. Antoniou$^{a,b}$
\footnote{e-mail address: nantonio@atlas.uoa.gr
\\
\begin{flushleft}
\normalsize CERN-TH/99-44\\February 1999
\end{flushleft}
          },
Y.F. Contoyiannis$^b$ and F.K. Diakonos$^b$ \\
\vspace{0.8cm}
{\it{$^a$ Theory Division, CERN, CH-1211 Geneva 23, Switzerland\\ 
$^b$ Department of Physics, University of Athens, GR-15771, Athens,
Greece}}\\
\end{center}
\vspace{1 cm}

\begin{abstract}
A critical point of second order, belonging to the
universality class of the $3d$ Ising model, has recently been advocated as a
strong candidate for the critical behaviour (at high temperatures) of 
QCD with non-zero quark masses. The implications of this conjecture are
investigated in the multiparticle environment of high-energy collisions.
A universal intermittency pattern of pion-density fluctuations is found, at
the critical point, and its association to the critical exponents is 
discussed. A Monte Carlo simulation of critical events, in heavy-ion 
collisions, reveals the detailed structure of these fluctuations, suggesting
a framework of (event-by-event) measurements in which the critical theory
of QCD may become falsifiable.
\end{abstract}

\setcounter{page}{0}

\newpage

The nature of the QCD critical behaviour, associated with chiral phase
transition, has gradually been better understood,
during the last few years, in
terms of the number of flavours and the primordial mass spectrum of the light
quarks. In an approximate world of two flavours $(u,d)$ with zero quark masses
$(m_u \approx m_d \approx 0)$, the high-temperature QCD phase transition
$(T_c \approx 150 \rm{MeV})$ is of second order [1--3]; it belongs to
the $O(4)$ universality class of a $3d$ Heisenberg magnet \cite{WR} and
the order parameter is associated with pion and sigma field condensates 
$\phi =(\vec{\pi}, \sigma)$. In a more realistic world with three flavours
$(u,d,s)$ and a finite mass $(m_s)$ of the strange quark, the system develops
a tricritical point $(m_s^*, T_c)$ at which the transition changes
from first $(m_s<m_s^*)$ to second order  $(m_s>m_s^*)$ \cite{CGr,WR}. In
the real world, however, with non-zero quark masses $(m_u,m_d \neq 0)$  
the second-order branch disappears from the phase diagram: the
corresponding transition
becomes a smooth crossover \cite{BJW} and the tricritical point is 
transformed into an endpoint of the first-order phase transition line
(critical point of second order). It has also been argued \cite{BR,HJSSV}
that the behaviour of the system near the QCD critical point $(T_c, \mu_c)$ 
belongs to the universality class of the Ising model in $3d$ (gas--liquid
transition); in this case, the order parameter is
associated with the
$\sigma$-field condensate alone,
$\sigma \sim \langle{\bar{\psi}}\psi \rangle$. The above
over-all picture of critical QCD has been advocated recently by a number
of authors [3--8] and despite the fact that one
cannot (yet) verify these ideas completely in a detailed {\it ab initio}
study on the lattice, owing to the involvement of a non-zero
chemical potential $\mu_c$ at the
critical point, this picture remains attractive, as it is based on 
rather general and universal properties of strong interactions, near  
criticality. In this Letter we claim that it can also be falsifiable, because 
it may lead to quantitative and sharp predictions for pion-density 
fluctuations, near the critical point. Such critical fluctuations can, 
in principle, be measured in relativistic heavy-ion 
collisions, by selecting rare events with strong and universal intermittency 
pattern \cite{BP} in momentum space.

To this end, consider the $3d$ effective action $\Gamma_c[\sigma]$, which
effectively describes the system of QCD at the critical point
$(T=T_c, \mu=\mu_c)$ :
\begin{equation}
\Gamma_c[\sigma]=T_c^{-1} \int d^3 \vec{x}\left[\frac{1}{2}(\nabla \sigma)^2
+ GT_c^4 (T_c^{-1} \sigma)^{\delta +1}\right]. 
\label{eq:action}
\end{equation}
In eq.(\ref{eq:action}) the macroscopic field $\sigma$ has a dimension 
$\sigma \sim (\rm{length})^{-1}$, $\delta$ is the isothermal
critical exponent,
and $G$, a dimensionless coupling, specifies the critical equation of state:
$\frac{\delta \Gamma_c}{\delta \phi} \sim G \phi^\delta$. Both parameters
$(G,\delta)$ are universal; in the $3d$ Ising class one may fix the
exponent at the mean field value $\delta \approx 5$, because of the smallness
of the anomalous dimension \cite{Tsyp}. The coupling $G$, for the same
universality class, has been estimated recently, in non-perturbative studies
of the critical equation of state, and its actual value is fixed, to a good
approximation, in the range $G \approx$ 1.5--2.5 \cite{Tsyp}.
Hence the only free parameter in the effective theory 
(\ref{eq:action}) is essentially
the critical temperature $T_c$, which determines the length
scale (in heavy-ion physics, $\beta_c=T_c^{-1} \sim$ 1--2 fm).

In order to test the theory (\ref{eq:action}) in the environment of a
high-energy collision, we have to first adapt
the longitudinal geometry to the
rapidity space ($\xi$) by introducing the proper time-scale ($\tau$) of the
collision $(A+A)$:
\begin{equation}
\Gamma_c[\sigma]=\frac{1}{C_A} \int d^2 \vec{x}_{\perp} d \xi \left[
\frac{1}{2}\left(\frac{\partial \sigma}{\partial \xi}\right)^2 +
\frac{\tau^2}{2}
(\nabla_{\perp} \sigma)^2 + G C_A^2 \beta_c^4 \sigma^6 \right],
\label{eq:acthi}
\end{equation}
where $C_A=\frac{\tau}{\beta_c}$. 
The effective field $\sigma(\xi, \vec{x}_{\perp})$, being macroscopic
(classical), is naturally associated with the density of $\sigma$-particles
in the $3d$ space $(\xi, \vec{x}_{\perp})$: $\sigma^2=
N_{\sigma}(\rm{volume})^{-1} \sim (\rm{length})^{-2}$. Technically,
this interpretation
follows from the requirement that the density matrix, associated
with the partition
function $Z=\int [{\cal{D}} \sigma] e^{-\Gamma_c}$ of the critical theory 
(\ref{eq:acthi}), be diagonal in the coherent-state representation of
the particles associated with the $\sigma$-field \cite{BSS}. As a
consequence, one finds for the average multiplicity
$\langle N_{\sigma} \rangle $ the form:
$\langle N_{\sigma} \rangle
=Z^{-1} \int [{\cal{D}} \sigma]e^{- \Gamma_c}
( \int d^2 \vec{x}_{\perp} d \xi~\sigma^2(\xi, \vec{x}_{\perp}))$, suggesting
that the appropriate order parameter for the study of the density
fluctuations of the isoscalar particles at the critical point, 
is $\sigma ^2(\xi, \vec{x}_{\perp})$ \cite{AKMDEPJC}.

In the cylindrical geometry of  the collision, the theory (\ref{eq:acthi}) 
can be solved by projecting out the effective action onto rapidity and
transverse space. Following the results in ref. \cite{ACDPPRL}, the 
critical fluctuations of the order parameter in these two spaces 
($1d$ and $2d$) are developed within clusters of size 
$\delta_c=\frac{\pi^{1/2} R_{\perp}}{2 \tau}$ and 
$r_{\perp c}=\frac{\pi^{3/2} R_{\perp}}{32}$ respectively ($R_{\perp}$ is the
transverse radius of the entire system at the critical point). These 
fluctuations are properly described (with the expected fractal dimension) by
the contribution to the partition function of instanton-like configurations,
which, within these clusters, are slowly varying functions and contribute
equally to the effective potential and to the derivative terms in 
eq. (\ref{eq:acthi}) \cite{ACDPPRL}. In this framework, the canonical
partition function, at $T=T_c$, for scalar particles ($\sigma$) in a
cylindrical volume $V \leq \pi \delta_c r_{\perp c}^2$, is written:
\begin{equation}
Z(N_{\sigma},V,T_c)=N_{\sigma}^{-1/2}
\exp\left[-\left(\frac{V_o}{V}\right)^2 N_{\sigma}^3\right],
\label{eq:partfun}
\end{equation}
where $V_o=\beta_c^2 \sqrt{2 G C_A}$.

The average multiplicity
$\langle N_{\sigma} \rangle$ of scalars,  in each $3d$ cluster,
as a function of the volume $V$, now follows from (\ref{eq:partfun}): 
\begin{equation}
\langle N_{\sigma} \rangle=
\frac{\Gamma(1/2)}{\Gamma(1/6)}(\frac{V}{V_o})^{2/3} ~~~~~~;
~~~~~~ V \gg V_o .
\label{eq:mult}
\end{equation}
Equation (\ref{eq:mult}) shows that the critical clusters are local, 
cylindrical fractals with a  minimal volume-scale $V=V_o$ where
self-similarity breaks down. The fractal dimensions in rapidity and transverse
space are $d_{F,1}=\frac{2}{3}$ and $d_{F,2}=\frac{4}{3}$, reflecting the
actual
value of the isothermal critical exponent $\delta \approx 5$ and also the
choice of the order parameter $\sigma^2$ ($d_{F,1}=
\frac{\delta -1}{\delta +1}$,
$d_{F,2}=\frac{2 (\delta-1)}{\delta +1}$). The corresponding density-density
correlation function obeys for each cluster a characteristic power-law:
\begin{equation}
\langle \rho(\xi,\vec{x}_{\perp})~\rho(0,0) \rangle =
\left(\frac{32}{\pi G C_A}\right)^{1/3}
\beta_c^{-2} \frac{\Gamma(1/2)}{\Gamma(1/6)}
\vert \xi \vert^{-1/3} \vert\frac{\vec{x}_{\perp}}{\beta_c} \vert^{-2/3} .
\label{eq:dedeco}
\end{equation}
Since the size of the critical clusters specifies also the correlation length
of the finite system \cite{ACDPPRL}, both in rapidity
($\delta_c=\frac{\sqrt{\pi} R_{\perp}}{2 \tau}
\left(\frac{G}{2} \right)^{-1/2}$)
and transverse space,
$(r_{\perp,c} = \frac{\pi^{3/2}}{32}(\frac{G}{2})^{-3/4} R_{\perp})$ we may
safely assume that
in the global volume $V_g = \pi R_{\perp}^2 \Delta$ ($\Delta$ is the total
rapidity available in the experiment) the critical clusters interact weakly,
so that they form an ideal classical gas with a Poisson multiplicity
distribution \cite{ADKLPLB}. The average multiplicity of clusters, in this
approximation, is
$\langle N_{cl} \rangle \approx \left( \frac{R_{\perp}}{r_{\perp,c}}
\right)^2 \left( \frac{\Delta}{\delta_c} \right)$ and, using
eq. (\ref{eq:mult}),
an upper bound of the excess multiplicity of pions produced when the system
passes through the critical point, can be extracted $(N_{\pi} \approx 2
N_{\sigma})$ $\frac{N_{\pi}}{\Delta} \leq \frac{16}{\sqrt{\pi}}
\left(\frac{R_{\perp}}{\beta_c} \right) \left(\frac{G}{2} \right)^{7/12}$.
On the other hand, the local
fractals, associated with the critical clusters finally build-up a global
fractal with the same dimension \cite{AnLes}. In fact, in the transverse
momentum space, the fractal $\tilde{F}_2$ is even stronger, since the
corresponding Fourier transform of the power-law (\ref{eq:dedeco}) gives
$\tilde{d}_{F,2}=\frac{2}{3}$ ($\tilde{d}_{F,2}=2-d_{F,2}$). On the basis
of these structures, the theory predicts a universal pattern of strong,
multidimensional intermittency (in rapidity, transverse momentum
and $3d$ momentum space)
of the form: $F^{(1)}_p \sim M^{\frac{p-1}{3}}$, $F^{(2)}_p \sim
(M^2)^{\frac{2 (p -1)}{3}}$, $F^{(3)}_p \sim (M^3)^{\frac{5 (p-1)}{9}}$
($p=2,3,.$) where $F_p$ is a scaled factorial
moment of order $p$ \cite{BP} as a function of $M$, the number of
one-dimensional cells in momentum space.
Other power-laws coming from the $1d$
projections of the fractal $\tilde{F}_2$ or from the Cartesian product of two
fractals in the cylindrical geometry of the system, can also be written
\cite{AKMDEPJC}. The predicted intermittency power-laws describe
the density fluctuations of scalars produced at the critical
point quantitatively, relating these fluctuations to the
critical exponents of the $3d$ universality class.

We have performed a Monte Carlo simulation of the critical events in Pb+Pb
collisions in the SPS ($\Delta \approx 6$), corresponding
to the choice of the
parameters: $T_c=120~\rm{MeV}, R_{\perp}=30~\rm{fm}, \tau_c=25~\rm{fm}$.
We have also fixed
the coupling $G$ at the value $G=2$ \cite{Tsyp}. In momentum space, the
isoscalar particles ($\sigma$) are grouped in critical clusters (cylindrical)
with appropriate power-law correlations, dictated by the Fourier transform
of eq. (\ref{eq:dedeco}) in transverse space:
$\langle \tilde{\rho}(\xi,\vec{p}_{\perp}) \tilde{\rho}(0,0) \rangle \sim
\vert \xi \vert^{-1/3} \vert \beta_c \vec{p}_{\perp} \vert^{-4/3}$. The
distribution of the clusters in rapidity (projected) corresponds to an ideal
$1d$ gas, whereas in transverse momentum
space the following extra constraints
are imposed: (a) the clusters (their centers) are exponentially
distributed according to the temperature $T_c$, $d n_c \sim
\exp(-\frac{p_{\perp,c}}{T_c})$ and (b) the total transverse momentum
carried by the clusters is zero, $\sum_{i} \vec{p}_{\perp,c}^{(i)}=0$.
Moreover the size of the critical clusters in transverse momentum space, as
determined by the minimum length scale in transverse configuration space, is
given by the expression $\vert \Delta \vec{p}_{\perp} \vert \approx
\frac{T_c}{2}\left(\frac{R_{\perp}}{\tau_c}\right)^{1/2}
\left(\frac{G}{2} \right)^{-3/8} \left(\frac{\pi^3}{c_A}\right)^{1/4}$ and
for the above set of parameters we estimate
$\vert \Delta \vec{p}_{\perp} \vert \approx 70~\rm{MeV}$. With these
ingredients and the overall constraints on the
formation of sigmas at $T=T_c$, a typical, simulated event is obtained with
characteristics shown in Figs.1--3: the overall picture is consistent with a
$3d$ cylindrical fractal in momentum space (Fig. 1), which is decomposed into
a $2d$ self-similar structure in transverse momentum space (Fig. 2)
and a $1d$
self-similar structure in rapidity space (Fig. 3). The multidimensional
intermittency pattern clearly reveals a linear spectrum of indices in each
case, consistent with the critical exponents of the theory \cite{ACDPPRL}.

In the process of a phase transition in thermal equilibrium, the fluctuating
system is very likely to get frozen immediately after crossing the critical
point of second order $(\mu_c, T_c)$ \cite{SRS}; therefore, the
fluctuations predicted at $T=T_c$ may enter the horizon of observation
(in the corresponding events) without any distortion ($T_f \approx T_c$). It
remains to guarantee that the decay of $\sigma$ may proceed near the
$\pi \pi$ threshold, rather rapidly transferring
the critical fluctuations found at $T=T_c$ automatically to the
$\pi^-$ spectra $(\sigma \to \pi^+ \pi^-)$ (Figs. 1--3).
The isoscalar $\sigma$ has a very small mass at
$T=T_f$; therefore, the system of sigmas remains frozen and stable
until the mass $m_{\sigma}$, as a function of temperature, rises to the
$\pi \pi$ threshold value, $m_{\sigma}=2 m_{\pi}$. If the corresponding time
scale $\tau_{th}$ is sufficiently long (in particular events), the expansion
rate $\vert \frac{\dot{T}}{T} \vert \leq \frac{1}{3 \tau_{th}}$, for
$\tau \geq \tau_{th}$, becomes very small with respect to the decay rate of
$\sigma$ (even near the $\pi \pi$ threshold), also because the
isoscalar is very strongly coupled to the $\pi \pi$ system. Under these
circumstances, the sigmas with $m_{\sigma} \stackrel{>}{\sim} 2 m_{\pi}$ have
enough time to decay, before their mass increases further, leading to charged
pions $(\pi^-)$ with the same fluctuation pattern in momentum space. We have
verified numerically this decay mechanism, near the $\pi \pi$ threshold,
in the case of the typical event shown in Figs. 1--3. We have used for the
parameters of the $\sigma$-meson in vacuum $(T=0)$ the average values
$m_{\sigma}^{(0)} \approx \Gamma^{(0)}_{\sigma} \approx 0.8~\rm{GeV}$ of the
$f_o(400--1200)$ $\pi \pi$ resonance \cite{MPL}. It turns out that, if
$\tau_{th} \gg 10~\rm{fm}$, the decay $\sigma \to \pi \pi$
proceeds near $\pi \pi$
threshold and therefore the intermittency pattern found at $T=T_c$ for sigmas
(Figs. 1--3) is also a prediction for the density fluctuations of the
abundant charged pions ($\pi^-$) associated with the critical point.

In the framework of heavy-ion physics, we have derived a pattern of local
dynamical fluctuations of the classical $\sigma$-field associated with the
critical point in QCD with non-zero quark masses. This critical point belongs
to the $3d$ Ising universality class, but the same pattern is valid in any
universality class associated with a $3d$ scalar theory. The reason is that,
in these theories, the anomalous dimension $\eta$ is very small \cite{Tsyp},
leading to the same (mean field) value of the exponent: $\delta \approx 5$.
Hence a similar pattern of fluctuations is expected, for instance, in the
O(4) $3d$ theory \cite{ADKLPLB}, which describes the chiral QCD phase
transition in the limit of zero quark masses \cite{WR}. At the level of
observed pions, however, these theories are expected to lead to different
fluctuation patterns since the in-medium evolution (and decay) of scalars
varies from one universality class to the other. In the case of real QCD
with non-zero quark masses, we have argued that  in a particular class of
events, in which the in-medium evolution of $\sigma$ is rather slow,
$\tau_{th} \gg \tau_c$, the fluctuation pattern found at $T=T_c$ (Figs. 1--3)
is inherited by the charged decay products of $\sigma$ near the $\pi \pi$
threshold ($\pi^-$) and manifests itself as a clear signature of the
quark-hadron phase transition in nuclear collisions.\\

We thank A.L.S. Angelis and C.G. Papadopoulos for their comments and
suggestions on the present work.
{}

\vspace*{1.0cm}

\begin{center}
\large \bf {Figure Captions}
\end{center}

\vspace*{0.7cm}
Figure 1: (a) The distribution of the $\sigma$-particles of a typical
critical event in $3d$ momentum space. The multiplicity of $\sigma$'s is 138.
(b) The corresponding ($3d$) factorial moments of order $p=2,3,4$ in
a log-log plot. A linear fit (dashed line) indicating the intermittency
pattern for this event is included.

\vspace*{0.7cm}
Figure 2: (a) The distribution of the $\sigma$-particles in
transverse-momentum space for the event of Fig. 1.
(b) The corresponding ($2d$) factorial moments of order $p=2,3,4$ in
a log-log plot. A linear fit (dashed line) indicating the resulting
intermittency pattern is also shown.

\vspace*{0.7cm}
Figure 3: (a) The distribution of the $\sigma$-particles in rapidity for the
event of Fig. 1.
(b) The rapidity factorial moments of order $p=2,3,4$ in
a log-log plot together with a linear fit (dashed line) indicating the
corresponding intermittency pattern.

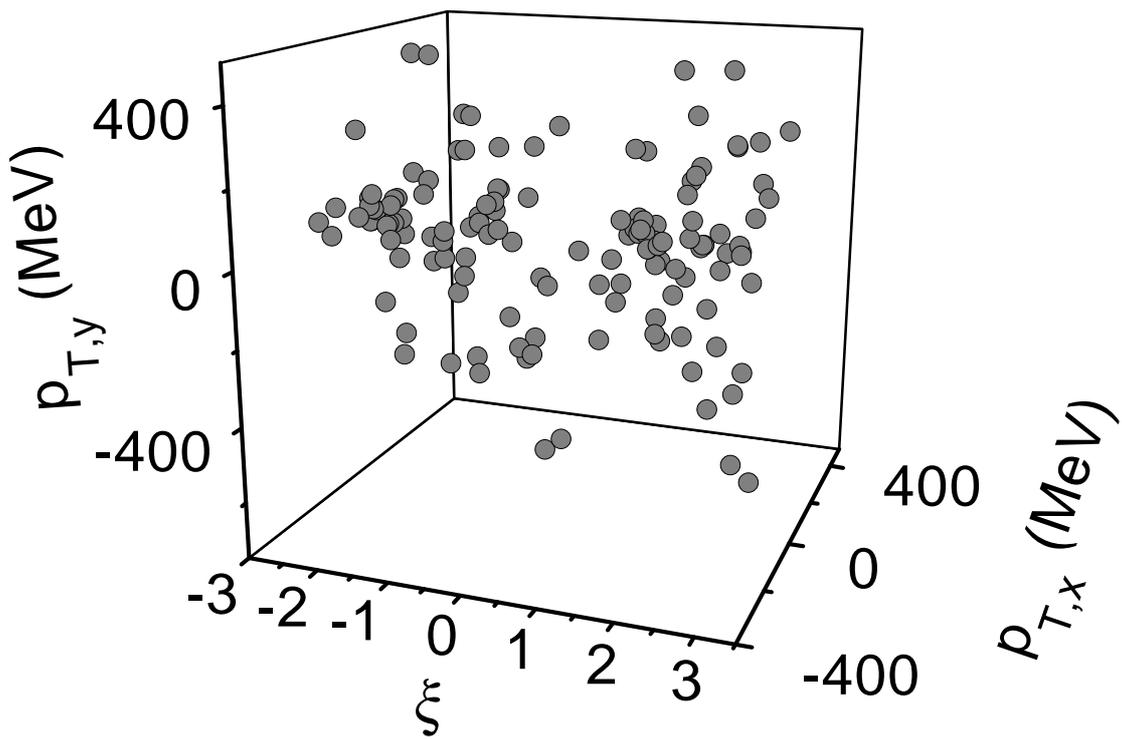
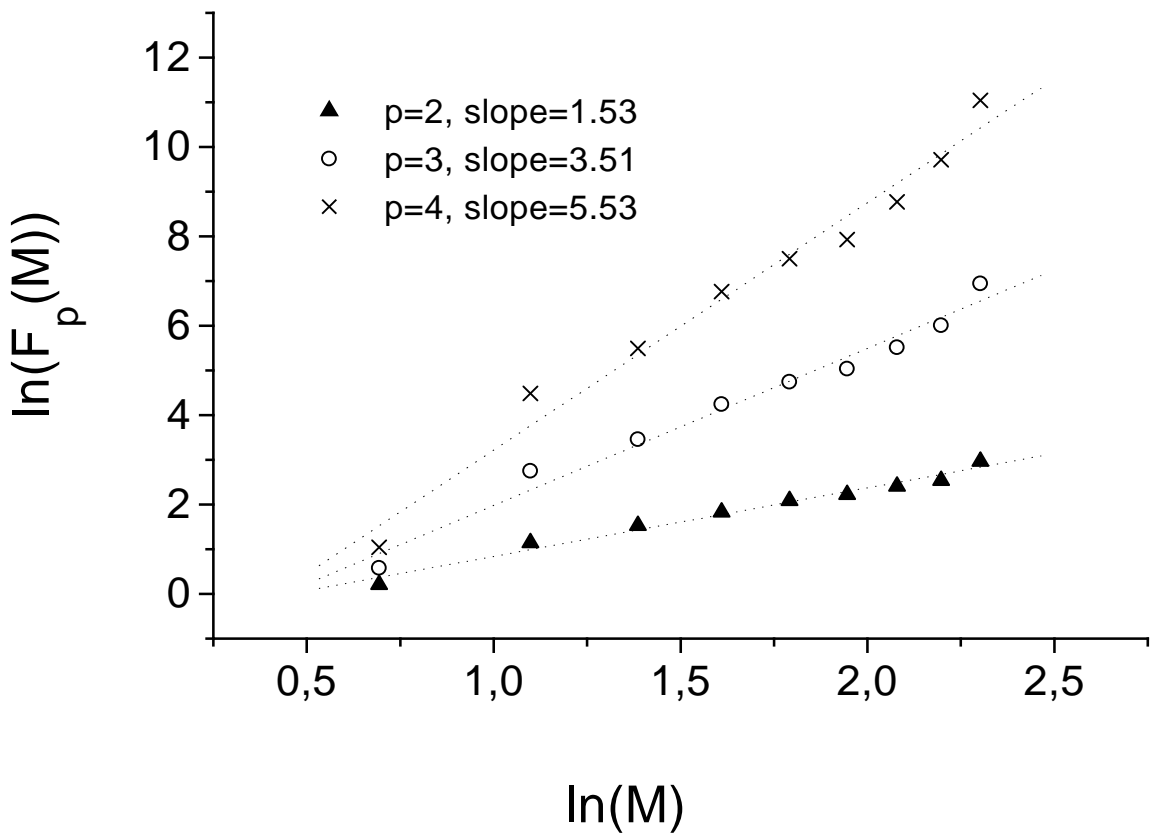
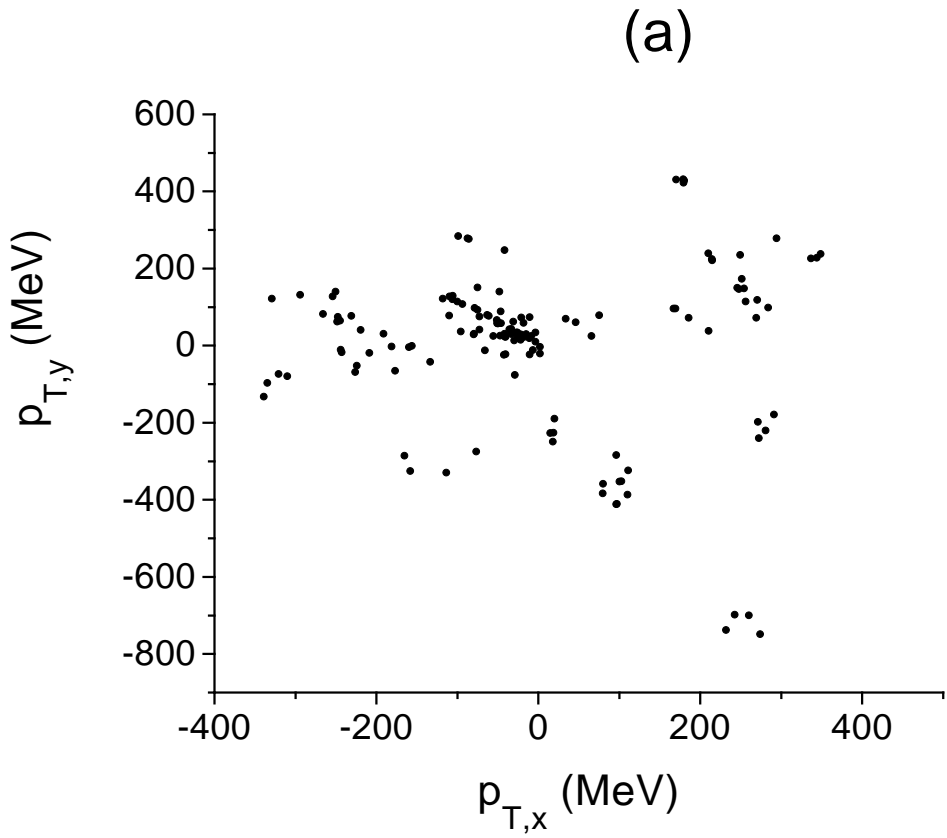
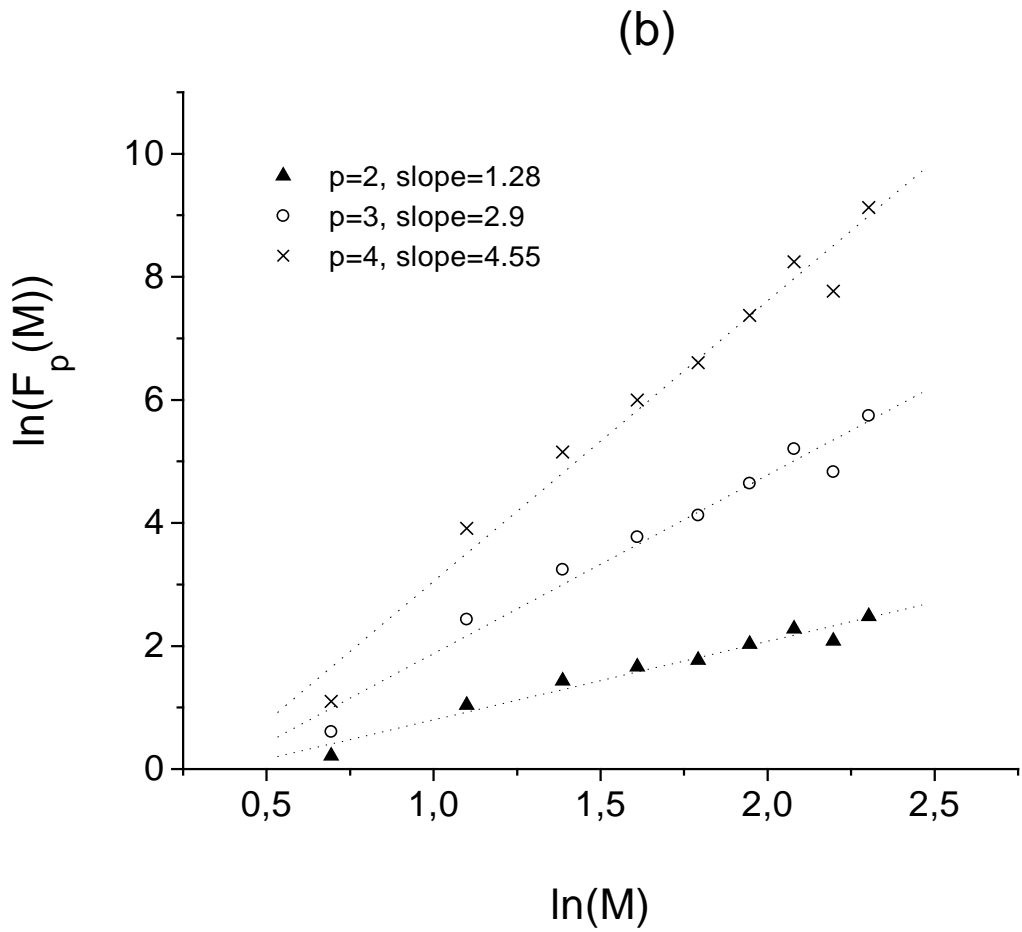
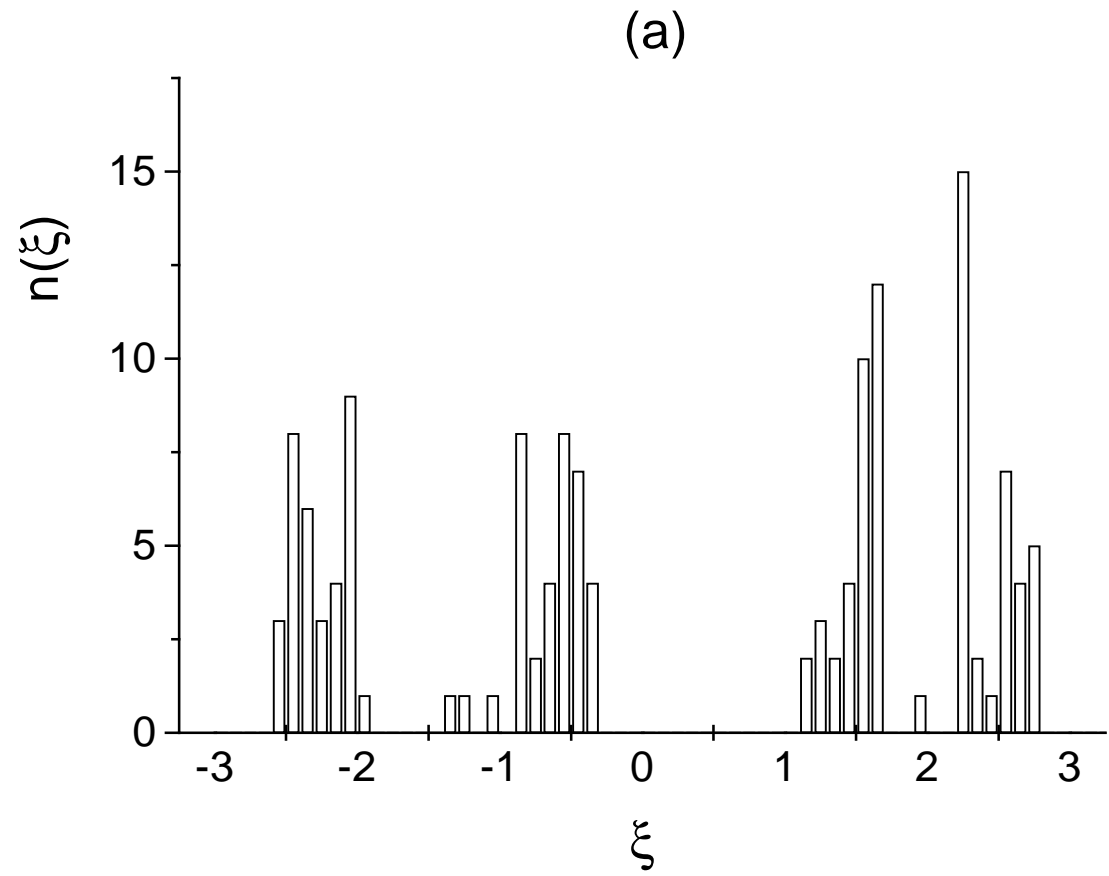
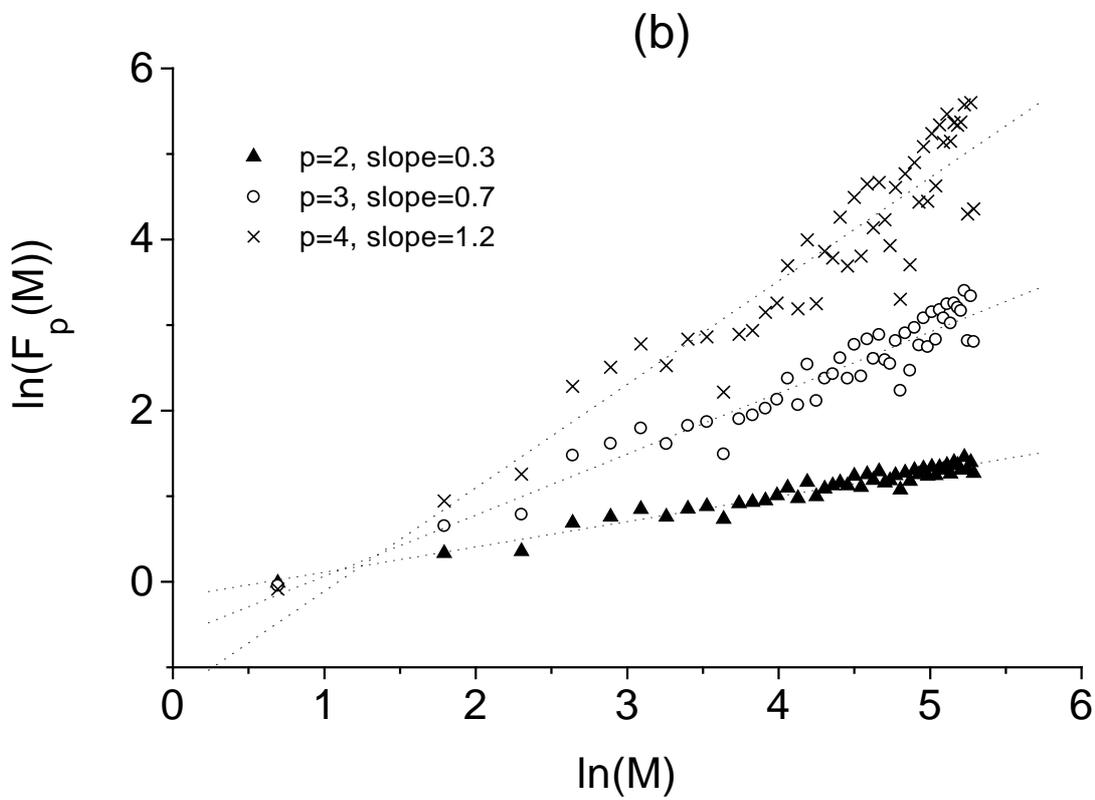
\end{document}